\documentclass[nofootinbib,aps,a4paper,10pt,letterpaper,superscriptaddress,twocolumn,showkeys,times,eqsecnum]{revtex4-1}
\usepackage{amsmath}
\usepackage{amsfonts}
\usepackage{lipsum}
\usepackage{times}

\usepackage{graphicx}
\usepackage{color}
\usepackage{braket}
\usepackage{dcolumn}
\usepackage{bm,url}
\usepackage[linktocpage]{hyperref}
\usepackage{subfigure}
\usepackage{amsfonts}
\usepackage{multirow}
\usepackage{booktabs}
\usepackage[left=2cm,right=2cm,top=2.5cm,bottom=2.5cm]{geometry}
\usepackage[usenames,dvipsnames,svgnames]{xcolor}  %% colored text
\usepackage{hyperref}   %% Needs to make hyper reference of any references or citations
\definecolor{oxfordblue}{rgb}{0.0, 0.13, 0.28}
\definecolor{burgundy}{rgb}{0.5, 0.0, 0.13}
\definecolor{darkolivegreen}{rgb}{0.33, 0.42, 0.18}
\definecolor{darkblue}{rgb}{0,0,0.5}
\definecolor{richcarmine}{rgb}{0.84, 0.0, 0.25}
\definecolor{darkblue}{rgb}{0,0,0.5}
\definecolor{bluer}{rgb}{0.00,0.50,0.75}{}
\hypersetup{colorlinks=true, citecolor=red, linkcolor=blue,
urlcolor = darkblue, filecolor=magenta}

\begin{document}

\newcommand\be{\begin{equation}}
\newcommand\ee{\end{equation}}
\newcommand\bea{\begin{eqnarray}}
\newcommand\eea{\end{eqnarray}}
\newcommand\bseq{\begin{subequations}} %solo con amsmath
\newcommand\eseq{\end{subequations}}
\newcommand\bcas{\begin{cases}}
\newcommand\ecas{\end{cases}}
\newcommand{\p}{\partial}
\newcommand{\f}{\frac}

\title{Primordial black hole in Lorentz-violating theories: Insights from Bumblebee gravity }
\author{Mohsen Khodadi}
\email{khodadi@kntu.ac.ir}
\affiliation{School of Physics, Institute for Research in Fundamental Sciences (IPM),	P. O. Box 19395-5531, Tehran, Iran}
\affiliation{School of Physics, Damghan University, Damghan 3671641167, Iran}
\affiliation{Center for Theoretical Physics, Khazar University, 41 Mehseti Str., AZ1096 Baku, Azerbaijan}

\author{Javad T. Firouzjaee}\email{firouzjaee@kntu.ac.ir}
\affiliation{Department of Physics, K. N. Toosi University of Technology,	P. O. Box 15875-4416, Tehran, Iran}

\date{\today}
%%%%%%%%%%%%%%%%%%%%%%%%%%%%%%%
\begin{abstract}
The Bumblebee gravity (BG) model, featuring spontaneous Lorentz symmetry breaking via a vector field non-minimally coupled to curvature, has been widely used to explore Lorentz-violating effects in cosmology. We investigate primordial black hole (PBH) formation within this framework, deriving the complete set of modified perturbation equations. We demonstrate that BG, sourced by a timelike vector field, introduces three distinct enhancements to PBH abundance—modified expansion history, suppressed collapse threshold, and amplified power spectrum—which together render PBHs viable dark matter candidates across the asteroid-mass window for modest Lorentz-violating couplings. However, a systematic analysis of the quadratic action reveals that these phenomenological consequences emerge from a theoretically pathological foundation. The vector sector exhibits an intrinsic ghost instability, while the requirement of a stable symmetry-breaking minimum simultaneously induces a tachyonic instability on timescales far below cosmological scales. The model thus suffers from a fundamental inconsistency: the conditions for cosmological viability and spontaneous symmetry breaking are mutually exclusive within the minimal Bumblebee framework. Our results illustrate both the notable power of Lorentz violation to influence early Universe observables and the necessity of a consistent theoretical foundation for such predictions.
	
\vspace{0.5cm}
\end{abstract}

\maketitle
\section{Introduction}
The nature of dark matter remains one of the most profound mysteries in modern cosmology, constituting approximately 85\% of the matter content of the Universe while eluding all non-gravitational detection efforts to date \cite{Bertone:2004pz, Bertone:2016nfn}. Among the myriad of proposed candidates, primordial black holes (PBHs) have experienced a remarkable renaissance over the past decade, evolving from a speculative curiosity to a theoretically well-motivated and observationally testable dark matter candidate \cite{Carr:1974nx, Carr:2020gox, Carr:2021bzv,Carr:2023tpt}. Formed from the gravitational collapse of overdense regions in the early Universe, PBHs naturally circumvent the requirement for new particle physics beyond the Standard Model and offer a purely gravitational explanation for dark matter \cite{Hawking:1971ei, Carr:1975qj}. The recent detection of gravitational waves from binary black hole mergers by LIGO/Virgo \cite{Abbott:2016blz} has reinvigorated interest in the possibility that these events originate from primordial populations, while simultaneously opening an unprecedented observational window into the high-energy physics operating during the first fractions of a second after the Big Bang \cite{Sasaki:2016jop, Bird:2016dcv, Clesse:2016vqa}.

The formation of PBHs is exquisitely sensitive to the statistical properties of primordial density perturbations and the gravitational dynamics governing their subsequent evolution \cite{Musco:2008hv, Musco:2012au, Harada:2013epa}. In the standard cosmological paradigm anchored by General Relativity and single-field slow-roll inflation, the amplitude of curvature perturbations on observable scales is constrained by cosmic microwave background (CMB) measurements to be approximately $\mathcal{P}_\zeta \sim 2.1\times 10^{-9}$ \cite{Planck:2018jri, Planck:2018vyg}. This value is far too small to generate cosmologically significant PBH abundances across most mass ranges \cite{Josan:2009qn, Carr:2009jm, Carr:2017jsz}. Consequently, substantial PBH production in General Relativity typically requires either a finely-tuned enhancement of the power spectrum on small scales—often realized via features in the inflationary potential \cite{Garcia-Bellido:1996mdl, Motohashi:2017kbs, Byrnes:2018txb} or multiple-field dynamics \cite{Bassett:2005xm}—or a violation of the radiation equation of state during the cosmic history that slows the Hubble expansion and facilitates collapse \cite{Carr:1975qj, Polnarev:2006aa, Afzal:2024xci}.

Recent studies have proposed that regular black holes could be interpreted as PBHs, thereby offering a potential explanation for dark matter and forging a novel link between the singularity problem and cosmological observables (see, e.g., Refs.~\cite{Davies:2024ysj,Calza:2024fzo,Calza:2024xdh,Calza:2025mwn,Loc:2025mzc})\footnote{This idea has not been without criticism \cite{Pacheco:2018mvs,Khodadi:2025icd}.}. In addition, it has recently been suggested that regular PBHs may also play a role in driving cosmic acceleration \cite{Dialektopoulos:2025mfz}.

This sensitivity to ultraviolet physics renders PBHs uniquely powerful probes of fundamental gravity theories beyond Einstein's formulation \cite{Sasaki:2018dmp, Green:2020jor, Ozsoy:2023ryl}. If PBHs are discovered and their mass function and abundance measured with precision, they will encode invaluable information about the gravitational dynamics operating at energy scales many orders of magnitude beyond terrestrial accelerator capabilities. Conversely, theoretical investigations into PBH formation within modified gravity frameworks can reveal both the phenomenological potential and the hidden pathologies of proposed extensions to General Relativity, providing essential guidance for model building and observational strategy.

One particularly intriguing class of modified gravity theories involves the explicit or spontaneous violation of Lorentz symmetry \cite{Kostelecky:2003fs, Mattingly:2005re, Liberati:2013xla}. Lorentz invariance stands as a cornerstone of both General Relativity and the Standard Model of particle physics, yet its violation is a generic prediction of various quantum gravity approaches.
%, including string theory \cite{Kostelecky:1988zi, Kostelecky:1989zf}, loop quantum gravity \cite{Gambini:1998it, Alfaro:2002wd}, and non-commutative geometries \cite{Carroll:2001ws, Mocioiu:2001wq}. 
The search for observable signatures of Lorentz violation has therefore emerged as a primary focus of modern fundamental physics, with extensive experimental programs spanning high-energy astrophysics \cite{Jacobson:2005bg, Liberati:2009pf}, precision laboratory measurements \cite{Kostelecky:2008ts, Hohensee:2011wt}, and cosmological observations \cite{Li:2008tma}.

Among the most widely studied phenomenological frameworks for Lorentz violation is the Bumblebee gravity (BG) model, in which Lorentz symmetry is spontaneously broken by the acquisition of a Vacuum Expectation Value (VEV) by a vector field coupled to curvature \cite{Kostelecky:2003fs, Bluhm:2004ep}. The model's relative simplicity, combined with its capacity to generate rich phenomenology while remaining technically tractable, has led to extensive investigations spanning black hole physics \cite{Bertolami:2005bh,Casana:2017jkc, Maluf:2020kgf,Khodadi:2021owg, Khodadi:2022pqh,Khodadi:2023yiw,Lambiase:2023zeo,AraujoFilho:2024iox,Mai:2024lgk,Pantig:2024lpg,Ding:2024qrf,Liu:2024axg,AraujoFilho:2025zaj}, neutron stars \cite{Ji:2024aeg, LIU:2024abe,Luo:2026oxw}, gravitational waves \cite{Liang:2022hxd,AouladLafkih:2025stw,Khodadi:2025wuw}, and cosmological dynamics \cite{Maluf:2021lwh,Khodadi:2022mzt,Khodadi:2023ezj,Nilsson:2023exc,Sarmah:2024xwx,Zhu:2024qcm,Reyes:2024hqi,vandeBruck:2025aaa,Nilsson:2025rvv,Siquieri:2026dby}. The model's defining feature—a non-minimal coupling $B^\mu B^\nu R_{\mu\nu}$ between the vector field and the Ricci tensor—introduces a coupling parameter $\xi$ that controls the strength of Lorentz-violating effects. When the vector field acquires a timelike/spacelike vacuum expectation value $b_0$, this coupling generates modifications to both the background expansion history and the evolution of cosmological perturbations, effects quantified by the dimensionless combination $\lambda \equiv \xi b_0^2$.

Despite its extensive phenomenological application, the Bumblebee model's theoretical consistency has long been a subject of concern. Early investigations identified potential issues with the model's constraint structure and the propagation of degrees of freedom \cite{Bluhm:2004ep, Bluhm:2007bd}, while subsequent work raised questions regarding the stability of cosmological solutions. The model's kinetic term for the vector field retains the Maxwell form $-\frac{1}{4}B_{\mu\nu}B^{\mu\nu}$. For timelike condensates, this choice is known to be problematic because the longitudinal mode of a massive vector field typically requires a Proca-type kinetic structure to avoid ghost instabilities. Furthermore, the non-minimal coupling $B^\mu B^\nu R_{\mu\nu}$ bears superficial resemblance to healthy couplings found in Horndeski and generalized Proca theories \cite{Horndeski:1974wa,Heisenberg:2014rta}. However, it lacks the special degeneracy structures that protect those theories from Ostrogradsky instabilities \cite{Langlois:2015cwa, Crisostomi:2016czh}. These concerns, however, have frequently been minimized or overlooked in phenomenological studies focused on observational consequences.

PBH formation provides an exceptionally stringent test of modified gravity theories precisely because it couples the background cosmological expansion, the generation and evolution of primordial perturbations, and the nonlinear dynamics of gravitational collapse. Any modification to gravity that survives at early times will imprint characteristic signatures across each of these regimes, and the consistency requirements linking them impose severe constraints on viable theoretical extensions \cite{Kawaguchi:2022nku,Nojiri:2024cal,Asfour:2025stv,Bousder:2025zfa}. The PBH formation process therefore offers a unique arena wherein the phenomenological promise and theoretical pathologies of models such as BG can be evaluated simultaneously and quantitatively.

In this work, we undertake a comprehensive investigation of PBH formation within the BG framework, pursuing two parallel objectives that are inextricably linked. Our first aim is to compute, for the first time, the complete set of observational predictions for PBH abundances emerging from the minimal Bumblebee model during inflation and the subsequent radiation-dominated era. This requires the systematic derivation of modified Friedmann equations incorporating the non-minimal coupling's backreaction, the full perturbation equations for scalar modes in the presence of the vector condensate, the quadratic action for curvature and vector perturbations in the uniform inflaton gauge, and the resulting modifications to the primordial power spectrum, the critical collapse threshold, and the PBH mass function. Through this analysis, we demonstrate that BG for modest Lorentz-violating couplings $\lambda$ and observationally consistent slow-roll parameters $\epsilon$, amplifies PBH abundances relative to General Relativity, rendering PBHs viable dark matter candidates across the asteroid-mass window $10^{17}\mathrm{g} \lesssim M \lesssim 10^{23}\mathrm{g}$ where observational constraints remain permissive \cite{Carr:2020gox}.

In parallel with the phenomenological calculation, we present a detailed assessment of the minimal Bumblebee model's theoretical consistency as a foundation for such predictions. By constructing the complete quadratic action for perturbations and systematically analyzing its structure, we identify two distinct pathologies inherent to the model's definition. First, the vector perturbation exhibits a ghost instability whose origin lies in the kinetic structure combined with the timelike vacuum expectation value. Second, the potential designed to spontaneously break Lorentz symmetry, while mathematically requiring a stable minimum, simultaneously generates an instability that disrupts the cosmological background on timescales far shorter than typical inflationary or radiation-dominated epochs. These issues appear to be inherent to the model's formulation rather than artifacts of gauge choice or perturbative approximations.

Our central thesis, developed throughout this work, is that these two pathologies—the ghost and the tachyon—are not separable issues but rather manifestations of a deeper incompatibility between the model's phenomenological ambitions and its theoretical structure. The same non-minimal coupling that generates the striking PBH enhancements also participates in the constraint equations that would, in a healthy theory, eliminate ghostly degrees of freedom, but fails to do so because the Maxwell kinetic term lacks the necessary degeneracy structure. The same potential that successfully enforces $B^\mu B_\mu = -b_0^2$ at the classical level inevitably produces tachyonic instabilities because the condition for a minimum and the condition for positive mass squared are one and the same. The extraordinary sensitivity of PBH abundances to the Lorentz-violating parameter $\lambda$ is not merely a phenomenological opportunity but a symptom—a direct signature of the tachyonic instability that renders the model inconsistent.

The structure of this paper reflects our twofold objectives. In Section \ref{Back}, we establish the modified background cosmology in BG, deriving the complete set of modified Friedmann equations and their solutions during radiation domination and slow-roll inflation. Section \ref{evolv} presents the linear perturbation theory for scalar modes, deriving the modified density contrast evolution equation, the effective Newton's constant, and the critical collapse threshold modification. In Section \ref{power}, we perform the complete quadratic action analysis in the uniform inflaton gauge, computing the kinetic, gradient, and mass matrices, identifying the ghost and tachyonic instabilities, and deriving the modified curvature power spectrum and sound speeds. Section \ref{pic} synthesizes these results into the full PBH abundance calculation, presenting numerical results across the asteroid-mass window for observationally consistent parameters and quantifying the enhancement factors.  In the Section \ref{con}, we presents our conclusions.

Throughout this work, we adopt units with $\hbar = c = k_B = 1$, the reduced Planck mass $M_p = (8\pi G)^{-1/2}$, and the metric signature $(-,+,+,+)$. Our perturbative calculations are performed to linear order in all perturbations and to leading order in the Lorentz-violating parameter $\lambda$ where appropriate, while retaining exact expressions where necessary for stability analysis.

\section{Modifying the background expansion history}\label{Back}

We begin with the minimal version of BG action \cite{Bluhm:2004ep,Bertolami:2005bh}
\be\label{act}
\begin{aligned}
	S = \int d^4x \sqrt{-g} \bigg[ & \frac{1}{2\kappa}(R + \xi B^\mu B^\nu R_{\mu\nu})  \\
	& - \frac{1}{4}B^{\mu\nu}B_{\mu\nu}  - V(B^\mu B_\mu \pm b^2) + \mathcal{L}_m \bigg],~~~\kappa \equiv 8\pi G
\end{aligned}
\ee
Where \(B_{\mu\nu} = \partial_\mu B_\nu - \partial_\nu B_\mu\), and \(V\) is a potential forcing spontaneous Lorentz violation: \(B^\mu B_\mu \pm b^2 = 0\). \(\mathcal{L}_m\) denotes the matter Lagrangian. In this action the bumblebee vector field couples non-minimally to curvature via dimensionless coupling parameter $\xi$.
 
The gravitational field equations come from \(\delta S/\delta g^{\mu\nu} = 0\). The variation gives
  \be
 \frac{1}{\sqrt{-g}}\frac{\delta S}{\delta g^{\mu\nu}} = \frac{1}{2\kappa}(G_{\mu\nu} + \xi K_{\mu\nu}) - \frac{1}{2}T_{\mu\nu}^B - \frac{1}{2}T_{\mu\nu}^m = 0
 \ee
  Where: \(G_{\mu\nu} = R_{\mu\nu} - \frac{1}{2}g_{\mu\nu}R\) is the Einstein tensor, \(T_{\mu\nu}^m\) is the matter energy-momentum tensor,  \(T_{\mu\nu}^B\) is the Bumblebee field energy-momentum tensor, and  \(K_{\mu\nu}\) is the additional term from the non-minimal coupling. Concerning the later i.e., \(K_{\mu\nu}\) tensor, the variation of the non-minimal term is
 \bea
\delta(\sqrt{-g} B^\alpha B^\beta R_{\alpha\beta}) = B^\alpha B^\beta \delta(\sqrt{-g} R_{\alpha\beta}) + \sqrt{-g} R_{\alpha\beta} \delta(B^\alpha B^\beta),
 \eea
where after some manipulation, we get
 \bea\label{K}
 \begin{aligned}
 	K_{\mu\nu} = &-\frac{1}{2}g_{\mu\nu}B^\alpha B^\beta R_{\alpha\beta} 
 	+ B_\mu B^\alpha R_{\alpha\nu} + B_\nu B^\alpha R_{\alpha\mu} \\
 	&+ \frac{1}{2}\nabla_\alpha \nabla_\mu (B^\alpha B_\nu) +
 	\frac{1}{2}\nabla_\alpha \nabla_\nu (B^\alpha B_\mu) \\
 	&- \frac{1}{2}\Box(B_\mu B_\nu) - \frac{1}{2}g_{\mu\nu}\nabla_\alpha \nabla_\beta (B^\alpha B^\beta)
 \end{aligned}
 \eea
The Bumblebee energy-momentum tensor, reads as
 \be
 T_{\mu\nu}^B = B_{\mu\alpha}B_\nu^{\ \alpha} - \frac{1}{4}g_{\mu\nu}B^{\alpha\beta}B_{\alpha\beta} + 2V' B_\mu B_\nu - g_{\mu\nu}V~,
 \ee
where \(V' = dV/dX\) with \(X = B^\mu B_\mu \pm b^2\).
  
 We adopt a flat FLRW metric:
 \be
 ds^2 = -dt^2 + a^2(t)(dx^2 + dy^2 + dz^2)
 \ee
 with timelike VEV Bumblebee field, 
 \be\label{BB}
 B_\mu = (b(t), 0, 0, 0)
 \ee
 obeying the constraint \(B^\mu B_\mu = -b(t)^2 = -b_0^2~~ ( \text{constant})\).
It is important to note that the timelike configuration (\ref{BB}) in a flat FLRW metric is spatially homogeneous by construction and consequently satisfies the cosmological principle at the background level. Although the field defines a preferred frame by pointing uniformly in the time direction everywhere, it maintains spatial isotropy since all spatial directions remain equivalent.
 
Now, let us compute components of $K_{\mu\nu}$ systematically.
 
\textbf{Time-time component \(K_{00}\):}
 \be\label{K00}
 \begin{aligned}
 	K_{00} = & -\frac{1}{2}g_{00}B^\alpha B^\beta R_{\alpha\beta} + 2 B_0 B^\alpha R_{\alpha 0} \\
 	& + \nabla_\alpha \nabla_0 (B^\alpha B_0) - \frac{1}{2}\Box(B_0 B_0) \\
 	& - \frac{1}{2}g_{00}\nabla_\alpha \nabla_\beta (B^\alpha B^\beta)
 \end{aligned}
 \ee
Since \(B_\mu = (b_0, 0, 0, 0)\) and \(b_0\) is constant so
 \(B^\alpha B^\beta R_{\alpha\beta} = B^0 B^0 R_{00} = b_0^2 R_{00} = -3b_0^2 \frac{\ddot{a}}{a}\), and \(B_0 B^\alpha R_{\alpha 0} = -b_0^2 R_{00} = 3b_0^2 \frac{\ddot{a}}{a}\), resulting in $ K_{00} = \frac{9}{2}b_0^2 \frac{\ddot{a}}{a}$. Also, the derivative terms vanish since \(b_0\) is constant and the field is homogeneous. 
 
\textbf{Space-space component \(K_{ij}\):}
\bea
\begin{aligned}
	K_{ij} = & -\frac{1}{2}g_{ij}B^\alpha B^\beta R_{\alpha\beta} + B_i B^\alpha R_{\alpha j} \\
	& + B_j B^\alpha R_{\alpha i} + \text{(derivative terms)}
\end{aligned}
\eea
The spatial components \(B_i = 0\), so many terms vanish:
 \be
 K_{ij} = -\frac{1}{2}g_{ij}(-3b_0^2 \frac{\ddot{a}}{a}) = \frac{3}{2}b_0^2 \frac{\ddot{a}}{a} g_{ij}
 \ee
Computing \(T_{\mu\nu}^B\) components: since \(B_\mu = (b_0, 0, 0, 0)\) is constant: \(B_{\mu\nu} = \partial_\mu B_\nu - \partial_\nu B_\mu = 0\), the potential enforces \(B^\mu B_\mu = -b_0^2\), so \(V = 0\) and \(V' \neq 0\) in general. Thus
 \be
 T_{\mu\nu}^B = 2V' B_\mu B_\nu - g_{\mu\nu}V = 2V' B_\mu B_\nu
 \ee
Specifically: \(T_{00}^B = 2V' B_0 B_0 = 2V' b_0^2\), and \(T_{ij}^B = 2V' B_i B_j = 0\). Now, we can derive the modified Friedmann equations.
 
Time-Time component (\(00\)) of the full field equation reads off
 \be
 G_{00} + \xi K_{00} = \kappa(T_{00}^B + T_{00}^m)
 \ee
Substituting \(G_{00} = 3H^2\), \( K_{00} = \frac{9}{2}b_0^2 \frac{\ddot{a}}{a}\), \(T_{00}^B = 2V' b_0^2\), and \(T_{00}^m = \rho_m\),
we get
 \be\label{tt}
 3H^2 + \xi\frac{9}{2}b_0^2 \frac{\ddot{a}}{a} = \kappa(2V' b_0^2 + \rho_m)
 \ee
For space-space component (\(ij\)) of  of the full field equation, we have
 \be
 G_{ij} + \xi K_{ij} = \kappa(T_{ij}^B + T_{ij}^m)
 \ee
Substituting \(G_{ij} = -a^2\delta_{ij}(3H^2 + 2\dot{H})\), \(K_{ij} = \frac{3}{2}b_0^2 \frac{\ddot{a}}{a} g_{ij} = \frac{3}{2}b_0^2 \frac{\ddot{a}}{a} a^2\delta_{ij}\),  \(T_{ij}^B = 0\), and \(T_{ij}^m = p_m g_{ij} = p_m a^2\delta_{ij}\), we come to 
\be\label{rr}
 -(3H^2 + 2\dot{H}) + \xi\frac{3}{2}b_0^2 \frac{\ddot{a}}{a} = \kappa p_m
 \ee
Besides, the Bumblebee field equation from \(\delta S/\delta B^\mu = 0\) gives:
 \be \label{b}
 \nabla^\mu B_{\mu\nu} = 2V' B_\nu - \frac{\xi}{\kappa} B^\mu R_{\mu\nu}
 \ee
For \(\nu = 0\) component: \(\nabla^\mu B_{\mu 0} = 0\) (since \(B_{\mu\nu} = 0\))
, and Eq. (\ref{b}) takes the following form
\be
2V' B_0 - \frac{\xi}{\kappa} B^\mu R_{\mu 0} = 0
\ee
By solving it in terms of $V'$, we have 
 \be
 V' = \frac{3\xi}{2\kappa} \frac{\ddot{a}}{a}
 \ee
Now, by taking \(V'\) into account of the time-time component (\ref{tt}), we arrive at
 \be
 3H^2 + \frac{3}{2}\xi b_0^2 \frac{\ddot{a}}{a} = \kappa\rho_m
 \ee
where by including \(\ddot{a}/a = \dot{H} + H^2\), it take the form as follow
 \be
 \left(3 + \frac{3}{2}\xi b_0^2\right)H^2 + \frac{3}{2}\xi b_0^2 \dot{H} = \kappa\rho_m
 \ee
As a result, by defining the parameter \(\lambda = \xi b_0^2\) characterizing  the Lorentz violation strength, the modified Friedmann equations become:
\bea
\begin{aligned}
		&\left(3 + \frac{3}{2}\lambda\right)H^2 + \frac{3}{2}\lambda \dot{H} = \kappa\rho_m\\
		&\left(-3 + \frac{3}{2}\lambda\right)H^2 + \left(-2 + \frac{3}{2}\lambda\right)\dot{H} = \kappa p_m
	\end{aligned}
\eea
By regarding small values of $\lambda$, the exact solution for the scale factor in Bumblebee gravity during radiation domination \(p_m = \frac{1}{3}\rho_m\) is 
\bea \label{scal}
&a(t) \approx a_0 t^{\frac{1}{2}-\frac{\lambda}{8} }\\
&H(t)\approx(\frac{1}{2}-\frac{\lambda}{8})\frac{1}{t}
\eea
The \(\dot{H}\) terms are most significant during rapid expansion changes.
Even small \(\lambda\) can significantly alter \(H(t)\) during the brief period of PBH formation, changing the horizon mass relation \(M_H(t)\).

The complete derivation shows that Bumblebee gravity modifies both the \(H^2\) and \(\dot{H}\) terms in the Friedmann equations, leading to a expansion history that differs from general relativity, particularly during periods of rapid expansion change - exactly the conditions relevant for PBH formation.

\subsection{Slow-roll parameters in BG} 
 We consider a canonical scalar field (the inflaton) with action:
 \be
 S_m = \int d^4x \sqrt{-g} \left[ -\frac{1}{2} g^{\mu\nu} \partial_\mu \phi \partial_\nu \phi - U(\phi) \right]
 \ee
 with the energy-momentum tensor is:
 \be
 T_{\mu\nu} = \partial_\mu \phi \partial_\nu \phi - g_{\mu\nu} \left[ \frac{1}{2} g^{\alpha\beta} \partial_\alpha \phi \partial_\beta \phi + U(\phi) \right]
 \ee
Given that in the background Solution, we deal with homogeneous field: \(\phi = \phi_0(t)\), and FLRW metric, then the background energy density and pressure are
 \be
 \rho_0 = \frac{1}{2} \dot{\phi}_0^2 + U(\phi_0), \quad p_0 = \frac{1}{2} \dot{\phi}_0^2 - U(\phi_0)
 \ee
 and the slow-roll parameter is
 \be
 \epsilon \equiv -\frac{\dot{H}}{H^2}
 \ee
By taking modified Friedman equations derived above, we arrive at
\bea\label{1}
&&\frac{1}{2}\dot{\phi}_0^2 = \frac{H^2}{2\kappa}\left[3\lambda + (2 - 3\lambda)\epsilon\right]
\\
&&U(\phi_0) = \frac{H^2}{\kappa}\left(3-\epsilon\right)\label{11}
\eea
For consistency, we require \(\frac{1}{2}\dot{\phi}_0^2 \geq 0\), and $U(\phi_0)>0$ which imposes constraints on \(\lambda\) (small values $\ll1$) and \(\epsilon\)
\bea
\epsilon\geq  -\frac{3\lambda}{2-3\lambda}~~~~\mbox{If}~~~ \lambda<\frac{2}{3}
\eea
It is clear that de Sitter space ($\epsilon\approx0,~~p\approx-\rho$) is physically allowed just for case of $\lambda>0$. This makes \(\lambda > 0\) (positive non-minimal coupling $\xi<0$) the more physically reasonable case for BG with canonical scalar field inflation. In other words, this suggests that for consistent slow-roll inflation in BG, we need very small values of $\lambda$ within allowed range $0\leq\lambda<\frac{2}{3}$, otherwise, the background is not de Sitter-like. As a result,
\be\label{BG}
H^2\approx H_{GR}^2(1-\frac{\lambda}{2})
\ee
A hint here is essential. Positive $\lambda$ (up to $\frac{2}{3}$) is the physically reasonable regime for slow-roll inflation in BG so that $\lambda<\frac{2}{3}$ comes from requiring the kinetic energy to remain positive during slow-roll.
This would mean the Lorentz-violating coupling $\xi=\frac{\lambda}{b_0^2}$ should be positive for consistency with inflation.

\section{Modifying density contrast evolution equation }\label{evolv}
 We serve the Newtonian gauge
  \be
 ds^2 = -(1 + 2\Phi)dt^2 + a^2(t)(1 - 2\Psi)\delta_{ij}dx^i dx^j
 \ee
with the energy-momentum tensor for a perfect fluid
 \be
 T^\mu_{\ \nu} = (\rho + p)u^\mu u_\nu + p\delta^\mu_{\ \nu}
 \ee
and perturbations
 \be
 \rho = \rho_0(t)(1 + \delta), \quad p = p_0(t) + \delta p, \quad u^\mu = (1 - \Phi, v^i)
 \ee
Here, \(v^i = \partial_i v\) is the velocity perturbation.
 
We must perturb the Bumblebee field equations. Let us begin with time-time Component (\(\mu\nu = 00\)). The perturbed 00-component of the field equations
includes the background:
 \be
 G_{00} = 3H^2, \quad T_{00} = \rho_0
 \ee
and perturbations:
\bea
 &&\delta G_{00} = -6H\dot{\Psi} + 6H^2\Phi + \frac{2}{a^2}\nabla^2\Psi
\\
 &&\delta T_{00} = \rho_0\delta + 2\rho_0\Phi
 \eea
For the \(K_{\mu\nu}\) term, we need \(\delta K_{00}\). From the definition (\ref{K}), for the background field with \(B_\mu = (b, 0, 0, 0)\), we have
 \bea
 K_{00} = -\frac{15}{2}b^2\frac{\ddot{a}}{a}
 \eea
 where the perturbation \(\delta K_{00}\) contains terms like \(b^2 \delta R_{00}\), \(b^2 \Phi \ddot{a}/a\), etc. After computation:
  \be
 \delta K_{00} = -3b^2\left[\ddot{\Psi} + 4H\dot{\Psi} - H\dot{\Phi} - (2\dot{H} + 3H^2)\Phi + \frac{1}{3a^2}\nabla^2(\Phi - 2\Psi)\right]
 \ee
 As a result, the perturbed 00-equation is:
 \bea
 \begin{aligned}
 	&\delta G_{00} + \xi \delta K_{00} = \kappa \delta T_{00}
 	\\
 	&-6H\dot{\Psi} + 6H^2\Phi + \frac{2}{a^2}\nabla^2\Psi \\
 	&\quad -3\xi b^2\bigg[\ddot{\Psi} + 4H\dot{\Psi} - H\dot{\Phi} - (2\dot{H} + 3H^2)\Phi \\
 	&\quad + \frac{1}{3a^2}\nabla^2(\Phi - 2\Psi)\bigg] = \kappa\rho_0(\delta + 2\Phi)\label{00}
 \end{aligned}
 \eea
By taking trace of space-space component, i.e., the trace of the ij-components gives another equation. After perturbation
 \be
 \delta G^i_{\ i} + \xi \delta K^i_{\ i} = \kappa \delta T^i_{\ i}
 \ee
 where the background is \(G^i_{\ i} = -6(\dot{H} + 2H^2)\), \(K^i_{\ i} = \frac{9}{2}b^2\frac{\ddot{a}}{a}\), and the perturbations read off
 \bea
 \begin{aligned}
 	\delta G^i_{\ i} = & 6\ddot{\Psi} + 18H\dot{\Psi} - 6H\dot{\Phi} - 12(\dot{H} + 2H^2)\Phi \\
 	& - \frac{2}{a^2}\nabla^2(\Phi - 2\Psi) \\[4pt]
 	\delta K^i_{\ i} = & 3b^2\bigg[\ddot{\Psi} + 7H\dot{\Psi} - H\dot{\Phi} - (2\dot{H} + 3H^2)\Phi \\
 	& + \frac{1}{3a^2}\nabla^2(\Phi - 2\Psi)\bigg] \\[4pt]
 	\delta T^i_{\ i} = & 3\delta p
 \end{aligned}
 \eea
As a result,
\be
\begin{aligned}
	6\ddot{\Psi} &+ 18H\dot{\Psi} - 6H\dot{\Phi} - 12(\dot{H} + 2H^2)\Phi \\
	&- \frac{2}{a^2}\nabla^2(\Phi - 2\Psi) \\
	&+ 3\lambda\bigg[\ddot{\Psi} + 7H\dot{\Psi} - H\dot{\Phi} - (2\dot{H} + 3H^2)\Phi \\
	&\qquad + \frac{1}{3a^2}\nabla^2(\Phi - 2\Psi)\bigg] = 3\kappa\delta p
\end{aligned}
\ee
For sub-horizon modes (\(k \gg aH\)), we can use the quasi-static approximation, i.e., the time derivatives are subdominant compared to spatial gradients.
In this limit, the $00$-equation becomes
  \be
 \frac{1}{a^2}\nabla^2\left[2(1 + \lambda)\Psi - \lambda \Phi\right] = \kappa\rho_0\delta
 \ee
 From the off-diagonal components, we get the relation between \(\Phi\) and \(\Psi\). In BG, generally \(\Phi \neq \Psi\). The ij-equation for \(i \neq j\) gives
 \bea
 &&\partial_i\partial_j(\Phi - \Psi) + \xi b^2 \partial_i\partial_j(\Phi - \Psi) = 0
\\
&& (1 + \xi b^2)(\Phi - \Psi) = 0
\eea
Thus $\Phi = \Psi$, meaning that the potentials are equal even in BG for scalar perturbations. With \(\Phi = \Psi\), the 00-equation becomes
   \be
 \frac{1}{a^2}\nabla^2\left[(2 + \lambda)\Psi\right] = \kappa\rho_0\delta
 \ee
Thus the modified Poisson equation is:
 \be
 \frac{1}{a^2}\nabla^2\Psi = 4\pi G_{\text{eff}}\rho_0\delta,~~~~~G_{\text{eff}} = \frac{G}{1 + \frac{\lambda}{2}}
 \ee

From the energy conservation condition i.e., \(\nabla_\mu T^{\mu 0} = 0\), we get the continuity equation:
\be
 \dot{\delta} + \frac{1}{a}(1 + w)\nabla^2 v - 3\dot{\Psi} = 0
 \ee
where \(w = p_0/\rho_0\).  From the momentum conservation condition, i.e., \(\nabla_\mu T^{\mu i} = 0\), we get the Euler equation
 \be
 \dot{v} + Hv + \frac{1}{a}\Phi + \frac{1}{a}\frac{\delta p}{\rho_0(1 + w)} = 0
 \ee
Now, we are in a right position to derive the growth equation. By taking time derivative of continuity equation:
  \be
 \ddot{\delta} + \frac{1}{a}(1 + w)\nabla^2 \dot{v} - \frac{\dot{a}}{a^2}(1 + w)\nabla^2 v - 3\ddot{\Psi} = 0
 \ee
 From Euler equation, we have
 \be
 \nabla^2 \dot{v} = -H\nabla^2 v - \frac{1}{a}\nabla^2\Phi - \frac{1}{a}\nabla^2\left(\frac{\delta p}{\rho_0(1 + w)}\right)
 \ee
 where after substitute into the time-derivative of continuity,
 we arrive at 
\be
 \ddot{\delta} - \frac{2H}{a}(1 + w)\nabla^2 v - \frac{1}{a^2}(1 + w)\nabla^2\Phi - \frac{1}{a^2}\nabla^2\left(\frac{\delta p}{\rho_0}\right) - 3\ddot{\Psi} = 0
 \ee
Now use the original continuity equation to eliminate \(v\)
 \be
 \frac{1}{a}\nabla^2 v = \frac{1}{1 + w}(3\dot{\Psi} - \dot{\delta})
 \ee
 Thus
 \be\label{del}
 \ddot{\delta} - 2H(3\dot{\Psi} - \dot{\delta}) - \frac{1}{a^2}(1 + w)\nabla^2\Phi - \frac{1}{a^2}\nabla^2\left(\frac{\delta p}{\rho_0}\right) - 3\ddot{\Psi} = 0
 \ee
 For sub-horizon modes, the \(\nabla^2\Phi\) term dominates. 
 
 By taking the adiabatic perturbations: \(\delta p = c_s^2 \rho_0 \delta\), with \(c_s^2 = \dot{p}_0/\dot{\rho}_0\), into the modified Poisson equation
  \be
 \frac{1}{a^2}\nabla^2\left(\frac{\delta p}{\rho_0}\right) = \frac{c_s^2}{a^2}\nabla^2\delta
 \ee
 Besides, for sub-horizon modes, we can also neglect the time derivatives of \(\Psi\) compared to spatial gradients in Eq. (\ref{del}), resulting in
\be
 \ddot{\delta} + 2H\dot{\delta} - \frac{c_s^2}{a^2}\nabla^2\delta - 4\pi G_{\text{eff}}\rho_0(1 + w)\delta = 0
 \ee
In Fourier space (\(\nabla^2 \to -k^2\)):
\be
\ddot{\delta}_k + 2H\dot{\delta}_k + \left(\frac{c_s^2 k^2}{a^2} - 4\pi G_{\text{eff}}\rho_0(1 + w)\right)\delta_k = 0~,
 \ee
 with the effective gravitational constant in BG i.e., $G_{\text{eff}} = \frac{2G}{2 + \lambda}$.
 For radiation domination (\(w = 1/3$, $c_s^2 = 1/3\), the final form reads off
  \be \label{gr}
 \ddot{\delta}_k + 2H\dot{\delta}_k + \left(\frac{k^2}{3a^2} - \frac{4H^2}{2+\lambda}\right)\delta_k = 0~,
 \ee
 where by relaxing BG parameter it recover rightly the standard expression.
The key feature of the density contrast evolution in BG is that modifies gravitational strength via $G_{\text{eff}} = \frac{2G}{2 + \lambda}$
 so that for \(\lambda > 0\), particularly worthy physical range $0\leq \lambda<\frac{2}{3}$ gravity becomes weaker gravity, indicating suppressed growth. The instability condition becomes
 \be
 \frac{k^2}{3a^2} < \frac{4H^2}{2+\lambda}
 \ee
 i.e., modes inside the modified Jeans length grow.
 
 This provides the foundation for studying PBH formation in BG, where both the perturbation growth and collapse threshold are modified compared to general relativity. 
 
\subsection{Solve of Eq. (\ref{gr})} 
 From our previous derivation of the correct scale factor during radiation domination in BG, the Hubble Parameter and density takes the following forms
\bea
&& H = \frac{\dot{a}}{a} = \frac{\alpha}{t}, \quad \dot{H} = -\frac{\alpha}{t^2}
 \\
&& \rho_0 = \frac{3H^2}{8\pi G} = \frac{3\alpha^2}{8\pi G t^2}
 \eea
where by substituting in density contrast evolution equation (\ref{gr}), we arrive at 
\be
\ddot{\delta}_k + \frac{2\alpha}{t}\dot{\delta}_k + \left(\frac{k^2}{3a_0^2 t^{2\alpha}} - \frac{4\alpha^2}{(2 + \lambda)t^2}\right)\delta_k = 0
 \ee
Now, we must be written it in conformal time. To do so, let us \(d\eta = dt/a\), so \(dt = a d\eta\). From \(a(t) = a_0 t^{\alpha}\), we have
 \be
 d\eta = \frac{dt}{a_0 t^{\alpha}} \quad \Rightarrow \quad \eta = \frac{t^{1-\alpha}}{a_0(1-\alpha)} + \text{const}
 \ee
Thus
 \be
 t \propto \eta^{\frac{1}{1-\alpha}}, \quad a \propto t^{\alpha} \propto \eta^{\frac{\alpha}{1-\alpha}}
 \ee
 and the conformal Hubble parameter reads off
 \be
 \mathcal{H} = \frac{a'}{a} = \frac{\alpha}{1-\alpha}\frac{1}{\eta}
 \ee
 After some straightforward manipulations the final form of (\ref{gr}) in terms of conformal time, is given by
  \be\label{grr}
 \delta_k'' + \frac{\alpha}{(1-\alpha)\eta}\delta_k' + \left[\frac{k^2}{3} - \frac{4\alpha^2}{(1-\alpha)^2(2+ \lambda)\eta^2}\right]\delta_k = 0
 \ee
Equation (\ref{grr}), in essence, is a Bessel-type equation as follows
 \be
 \delta_k'' + \frac{p}{\eta}\delta_k' + \left[k^2 c_s^2 - \frac{\nu^2 - \frac{1}{4}}{\eta^2}\right]\delta_k = 0
 \ee
 where \(p = \dfrac{\alpha}{1-\alpha}\), \(c_s^2 = \dfrac{1}{3}\), and $\alpha\approx \frac{1}{2}-\frac{\lambda}{8}$ (from (\ref{scal})) with
 \be
 \nu = \sqrt{\frac{1}{4} + \frac{4p^2}{2 + \lambda}}
 \ee
 The general solution is:
  \be
 \delta_k(\eta) = \eta^{\frac{1-p}{2}}\left[C_1 J_\nu(k c_s \eta) + C_2 Y_\nu(k c_s \eta)\right]
 \ee
 where \(J_\nu\) and \(Y_\nu\) are Bessel functions. Now, let us consider the asymptotic behavior of above solution. 
 
 For sub-horizon limit, i.e. \(k c_s \eta \gg 1\), we have
 \be
 \begin{aligned}
 	\delta_k(\eta) \approx \sqrt{\frac{2}{\pi k c_s \eta}} \eta^{\frac{1-p}{2}} \Bigg[ & C_1 \cos\left(k c_s \eta - \frac{\nu\pi}{2} - \frac{\pi}{4}\right) \\
 	& + C_2 \sin\left(k c_s \eta - \frac{\nu\pi}{2} - \frac{\pi}{4}\right) \Bigg]
 \end{aligned}
 \ee
 Namely, oscillations with amplitude \(\propto \eta^{\frac{1-p}{2} - \frac{1}{2}} = \eta^{-\frac{p}{2}}\). 
 
 For super-horizon limit, i.e., \(k c_s \eta \ll 1\), using \(J_\nu(x) \approx (x/2)^\nu/\Gamma(\nu+1)\) and \(Y_\nu(x) \approx -\Gamma(\nu)(x/2)^{-\nu}/\pi\), we obtain
 \be \label{tag{3.44}}
 \delta_k(\eta) \approx \eta^{\frac{1-p}{2}} \left[ \tilde{C}_1 (k\eta)^\nu + \tilde{C}_2 (k\eta)^{-\nu} \right].
  \ee
For adiabatic initial conditions, the growing mode corresponds to the particular solution driven by the \(k^2\) term, yielding
 \be \label{tag{3.45}}
 \delta_k(\eta) \propto k^2 \eta^2.
 \ee
  Since \(a \propto \eta^p\), we have \(\eta^2 \propto a^{2/p}\). With \(p \approx 1 - \lambda/2\) for small \(\lambda\),
 \be
 \frac{2}{p} \approx 2\left(1 + \frac{\lambda}{2}\right) = 2 + \lambda.
 \ee
Thus, the superhorizon growth in terms of the scale factor is
 \be \label{tag{3.46}}
 \delta_k(a) \propto k^2 a^{2+\lambda}.
  \ee
In the GR limit (\(\lambda = 0\)), we recover \(\delta_k \propto a^2\), as required.  It means that BG enhances the growth of density perturbations on superhorizon scales for \(\lambda > 0\) since the superhorizon growth exponent is $n \equiv \frac{d \ln \delta}{d \ln a} \approx 2 + \lambda$. This tiny deviation from scale invariant is important since the rate of super-horizon growth directly determines the amplitude of perturbations at horizon entry, which is the primary factor for PBH formation.

\subsection{Spherical Collapse with Modified Growth}
The super-horizon growth expression (\ref{tag{3.46}}) in terms of scale factor reads off
\be
\delta(a) \propto a^{n}, \quad n = 2+\lambda
\ee
In the spherical collapse model, the critical density contrast \(\delta_c\) is related to the linear extrapolation of the initial perturbation. If a perturbation has initial amplitude \(\delta_i\) when the scale factor is \(a_i\), then by the time of horizon crossing (\(a_h\)), it grows to
\be
\delta_h = \delta_i \left( \frac{a_h}{a_i} \right)^n
\ee
The condition for collapse is when this linearly extrapolated value reaches \(\delta_c\). In the underlying BG the growth exponent modifies as \(n = 2 +\lambda\) so that by setting \(\lambda = 0\) then recovers GR limit, i.e., \(n = 2\). As a result, $\frac{\delta_h^{\text{BG}}}{\delta_h^{\text{GR}}} = \left( \frac{a_h}{a_i} \right)^{\lambda}$, meaning that for the same $\delta_i$, the density contrast at horizon crossing is larger in BG if $\lambda>0$.

The critical threshold $\delta_c$ is the value of $\delta_h$ at horizon crossing required for collapse. If $\delta_h^{\text{BG}}>\delta_h^{\text{GR}}$
for the same initial perturbation, then a smaller initial perturbation is needed in BG to reach the same $\delta_c$.

For a fixed horizon-crossing amplitude \(\delta_h\), the required initial amplitude is
\be
\delta_i = \delta_h \left( \frac{a_i}{a_h} \right)^n
\ee
The ratio of initial amplitudes in BG vs GR is
\be
\frac{\delta_i^{\text{BG}}}{\delta_i^{\text{GR}}} = \left( \frac{a_i}{a_h} \right)^{n_{BG} - n_{\text{GR}}} 
\ee
Since \(\delta_c\) is essentially this initial amplitude required for collapse:
\be
\frac{\delta_c^{\text{BG}}}{\delta_c^{\text{GR}}} \approx \left( \frac{a_h}{a_i} \right)^{-\lambda}
\ee
Typically \(a_h/a_i \sim e^{N}\) where \(N \) e-folds for PBH scales.
As a result,
the critical density contrast in BG is
\be
\delta_c^{\text{BG}} \approx \delta_c^{\text{GR}} \exp\left[ -\lambda N \right]
\ee
where \(\delta_c^{\text{GR}} \approx 0.414\) \cite{Harada:2013epa,Musco:2012au}, and \(N\) is the number of e-folds between the time the perturbation is generated and horizon crossing. For $\lambda>0$, the collapse is easier in BG, making enhance PBH formation because the critical threshold is smaller, i.e., $\delta_c^{\text{BG}}<\delta_c^{\text{GR}}$.

\section{Modifying power spectrum $\mathcal{P}_{\mathcal{R}}(k)$ }\label{power}
Let us compute the power spectrum \(\mathcal{P}_\mathcal{R}(k)\) in BG, focusing explicitly on the instability mechanism. This is where the most significant effect occurs.

We work in uniform inflaton gauge (also often called the comoving gauge or unitary gauge for the inflaton), where the inflaton perturbation \(\delta\phi = 0\) and the spatial metric is perturbed as
\be
g_{ij} = a^2(t)e^{2\zeta}(\delta_{ij} + \gamma_{ij}), \quad \partial_i \gamma_{ij} = \gamma_{ii} = 0
\ee
The ADM metric is
\be
ds^2 = -N^2 dt^2 + \gamma_{ij}(dx^i + N^i dt)(dx^j + N^j dt)
\ee
with
\be
N = 1 + \alpha, \quad N_i = \partial_i \psi, \quad \gamma_{ij} = a^2 e^{2\zeta} \delta_{ij}
\ee
The Bumblebee field has a timelike VEV plus perturbations:
\be
B_0 = b(t) + \delta b_0, \quad B_i = \partial_i \delta b
\ee
The potential enforces the constraint \(B^\mu B_\mu = -b^2\) at the minimum, so \(V(X) = 0\) and \(V'(X) \neq 0\) in general. The metric perturbations $\alpha$, $\psi$, and $\delta b_0$, become constrained variables (they are determined by solving constraint equations).

To do so, we substitute the perturbed metric and fields into the action (\ref{act}) and expand to quadratic order. The Lagrangian density becomes
\be
\mathcal{L}^{(2)} = \mathcal{L}_{EH}^{(2)} + \mathcal{L}_{NM}^{(2)} + \mathcal{L}_{B}^{(2)} + \mathcal{L}_{V}^{(2)} + \mathcal{L}_{m}^{(2)}
\ee
where the terms correspond to Einstein-Hilbert, non-minimal coupling, Bumblebee kinetic, potential, and matter contributions.
The variables \(\alpha\), \(\psi\), and \(\delta b_0\) appear without time derivatives in the quadratic action. Their equations of motion are constraints.

	\subsection{Quadratic action}

The full quadratic action in uniform inflaton gauge is:
\begin{equation}
	S^{(2)} = \int d^4x \, \left( \mathcal{L}_{EH}^{(2)} + \mathcal{L}_{NM}^{(2)} + \mathcal{L}_{B,\mathrm{kin}}^{(2)} + \mathcal{L}_V^{(2)} + \mathcal{L}_m^{(2)} \right),
\end{equation}
with the following Lagrangian density expressions:
\subsubsection*{Einstein-Hilbert term}
\begin{equation}
	\begin{aligned}
		\mathcal{L}_{EH}^{(2)} = M_p^2 a^3 \bigg[ & -6H^2\alpha^2 - 6H\alpha\dot{\zeta} + 2H\alpha\frac{\nabla^2\psi}{a^2} \\
		& + 2\frac{(\nabla\zeta)^2}{a^2} + 3\dot{\zeta}^2 - 2\dot{\zeta}\frac{\nabla^2\psi}{a^2} \bigg].
	\end{aligned}
\end{equation}

\subsubsection*{Non-minimal coupling term}
With $\dot{b}=0$ and $\lambda = \xi b_0^2$:
\begin{equation}
	\begin{aligned}
		\mathcal{L}_{NM}^{(2)} = \xi M_p^2 a^3 \bigg[ & -3b_0^2 H^2 \alpha^2 - 3b_0^2 H \alpha \dot{\zeta} + b_0^2 H \alpha \frac{\nabla^2\psi}{a^2} \\
		& - 3H^2 b_0 \alpha \delta b_0 + \cdots \bigg].
	\end{aligned}
\end{equation}

\subsubsection*{Bumblebee kinetic term}
\begin{equation}
	B_{\mu\nu} = \partial_\mu B_\nu - \partial_\nu B_\mu, \quad B_{0i} = \partial_i (\dot{\delta}b - \delta b_0), \quad B_{ij}=0.
\end{equation}
\begin{equation}
	B^{\mu\nu}B_{\mu\nu} = 2g^{00}g^{ij} B_{0i}B_{0j} = 2(-N^{-2}) a^{-2} e^{-2\zeta} (\partial(\dot{\delta}b - \delta b_0))^2.
\end{equation}
Expanding $N^{-2} \approx 1 - 2\alpha$, $e^{-2\zeta} \approx 1 - 2\zeta$, and $\sqrt{-g} = a^3 (1 + \alpha + 3\zeta)$:
\begin{equation}
 \mathcal{L}_{B,\mathrm{kin}}^{(2)} = -\frac12 a (1 + \alpha + 3\zeta) \left[ (\partial\dot{\delta}b)^2 - 2 \partial\dot{\delta}b \cdot \partial\delta b_0 + (\partial\delta b_0)^2 \right] .
\end{equation}
There is no $\dot{\delta b}_0^2$ term, meaning that $\delta b_0$ is non-dynamical. The coefficient of $(\partial\dot{\delta}b)^2$ is negative, representing a ghost instability. The field $\delta b$ has a wrong-sign kinetic term even in flat space. This ghost is not a gauge artifact; it is a physical pathology of the Bumblebee model.

\subsubsection*{Potential term}
\begin{equation}
	\mathcal{L}_V^{(2)} = -a^3 \left[ V'(0) \delta^{(2)}X + \frac12 V''(0) (\delta X)^2 \right],
\end{equation}
\begin{equation}
	\delta X = 2b_0 \delta b_0 + 2\alpha b_0^2 + a^{-2}(\partial\delta b)^2.
\end{equation}
This gives mass terms and mixings, but no time derivatives.

\subsubsection*{Matter term (Inflaton)}
\begin{equation} \label{matter}
	\begin{aligned}
		\mathcal{L}_m^{(2)} = a^3 \bigg[ & \left( \frac12 \dot{\phi}_0^2 + U \right)(3\alpha\zeta + \tfrac92 \zeta^2) \\
		& - \dot{\phi}_0^2 (\alpha^2 + 3\alpha\zeta) + \tfrac12 \dot{\phi}_0^2 \alpha^2 \bigg],
	\end{aligned}
\end{equation}
with $\dot{\phi}_0^2$ and $U$ given by the background (\ref{1}, and \ref{11}). No dependence on $\delta b, \delta b_0$.

\subsection{Constraint equations}
The underlying system has four equations from varying the action with respect to: $\alpha$ (Hamiltonian constraint), $\psi$ (Momentum constraint), $\delta b_0$ (Bumblebee temporal constraint), and $\delta b$ (Bumblebee spatial constraint) — the dynamical equation

\subsubsection{Momentum Constraint ($\delta S/\delta\psi = 0$)}

\begin{align}
	\frac{\delta\mathcal{L}_{EH}}{\delta\psi} &= M_p^2 a^3 \left[ 2H\frac{\nabla^2\alpha}{a^2} - 2\frac{\nabla^2\dot{\zeta}}{a^2} \right], \\
	\frac{\delta\mathcal{L}_{NM}}{\delta\psi} &= \xi M_p^2 a^3 \left[ b_0^2 H\frac{\nabla^2\alpha}{a^2} - b_0^2 \frac{\nabla^2\dot{\zeta}}{a^2} \right].
\end{align}
Summing:
\begin{equation}
	M_p^2 a^3 (1+\lambda) \left[ 2H\frac{\nabla^2\alpha}{a^2} - 2\frac{\nabla^2\dot{\zeta}}{a^2} \right] = 0.
\end{equation}
Cancelling the non-zero factor $2M_p^2 a^3 (1+\lambda) \nabla^2/a^2$:
\begin{equation}
 H\alpha - \dot{\zeta} = 0 \quad \Rightarrow \quad \alpha = \frac{\dot{\zeta}}{H}.
\end{equation}
This is exact at linear order, no approximations.

\subsubsection{Hamiltonian constraint ($\delta S/\delta\alpha = 0$)}

The full Hamiltonian constraint is lengthy. For subhorizon modes ($k \gg aH$), the $\nabla^2\psi$ terms dominate. Using the standard solution for $\psi$ from the momentum constraint and spatial gauge condition
\begin{equation}
	\frac{\nabla^2\psi}{a^2} \approx -3\dot{\zeta} - \frac92 H\zeta + \mathcal{O}(\epsilon, \lambda).
\end{equation}
Substituting into the EH+NM part and keeping leading terms:
\begin{equation}
	\begin{aligned}
		& M_p^2 a^3 (1+\lambda) \left[ -12H^2\alpha - 6H\dot{\zeta} + 2H\frac{\nabla^2\psi}{a^2} \right] \\
		& \qquad \approx M_p^2 a^3 (1+\lambda) \left[ -12H^2\alpha - 12H\dot{\zeta} - 9H^2\zeta \right].
	\end{aligned}
\end{equation}
Adding matter and potential terms gives a relation between $\alpha, \dot{\zeta}, \zeta, \delta b_0$. This determines $\psi$ and $\delta b_0$, not $\alpha$ — $\alpha$ is already fixed by the momentum constraint.
In other words, the momentum constraint fixes $\alpha = \dot{\zeta}/H$. The Hamiltonian constraint then determines $\psi$ and $\delta b_0$ in terms of $\zeta$ and its derivatives. This is exactly the standard ADM procedure.

\subsubsection{Bumblebee temporal Constraint ($\delta S/\delta(\delta b_0) = 0$)}

\begin{align}
	\frac{\delta\mathcal{L}_{NM}}{\delta(\delta b_0)} &= \xi M_p^2 a^3 \left[ -3H^2 b_0 \alpha - b_0 \frac{\nabla^2\alpha}{a^2} \right], \\
	\frac{\delta\mathcal{L}_{B,\mathrm{kin}}}{\delta(\delta b_0)} &= a \frac{\nabla^2}{a^2} (\dot{\delta}b - \delta b_0), \\
	\frac{\delta\mathcal{L}_V}{\delta(\delta b_0)} &= 2a^3 b_0 V'(0) - 4a^3 b_0^2 V''(0) \delta b_0.
\end{align}
The background term $2a^3 b_0 V'(0)$ cancels with the background equation of motion. Linearizing:
\begin{equation}
\frac{\nabla^2}{a^2} (\dot{\delta}b - \delta b_0) - 3\xi b_0 H^2 \alpha - \xi b_0 \frac{\nabla^2\alpha}{a^2} - 4b_0^2 V''(0) \delta b_0 = 0.
\end{equation}
For \textbf{subhorizon modes} ($k \gg aH$), the $\nabla^2$ terms dominate:
\begin{equation}
	\delta b_0 = \dot{\delta}b - \xi b_0 \alpha.
\end{equation}
Using $\alpha = \dot{\zeta}/H$ from the momentum constraint:
\begin{equation}
\delta b_0 = \dot{\delta}b - \frac{\xi b_0}{H} \dot{\zeta}.
\end{equation}
This is the form used in the reduced action.

\subsubsection{Bumblebee spatial constraint ($\delta S/\delta(\delta b) = 0$)}

This is the dynamical equation of motion for $\delta b$, not a constraint. Varying $\mathcal{L}_{B,\mathrm{kin}}$:
\begin{align}
	\frac{\delta\mathcal{L}_{B,\mathrm{kin}}}{\delta(\delta b)} &= \frac{d}{dt} \left[ a (1+\alpha+3\zeta) \frac{\nabla^2}{a^2} (\dot{\delta}b - \delta b_0) \right] \\
	&\quad + \partial_i \left[ a (1+\alpha+3\zeta) \partial^i (\dot{\delta}b - \delta b_0) \right] + \cdots
\end{align}
Contributions from $\mathcal{L}_{NM}$ and $\mathcal{L}_V$ give mass and mixing terms. The full linearized equation is:
\begin{equation}
	\begin{aligned}
		\frac{d}{dt} \left[ a \frac{\nabla^2}{a^2} (\dot{\delta}b - \delta b_0) \right] &+ \frac{\nabla^2}{a} (\dot{\delta}b - \delta b_0) \\
		&+ \left( \frac{k^2}{a^2} + m_{\delta b}^2 \right) \delta b + \text{mixing with } \zeta = 0.
	\end{aligned}
\end{equation}
This is a second-order differential equation in time for $\delta b$. It propagates one degree of freedom, confirming our DOF count.

\subsection{Kinetic matrix and ghost analysis}

We now derive the full kinetic matrix for the dynamical fields \(\zeta\) and \(\delta b\) after eliminating the constrained variables \(\alpha, \psi, \delta b_0\). The quadratic action in uniform inflaton gauge reduces to:

\be\label{actt}
\begin{aligned}
	S^{(2)} = \frac{1}{2} \int d^4x & \left\{ a^3 \left[ \mathcal{G}_{11} \dot{\zeta}^2 + 2\mathcal{G}_{12} \dot{\zeta} \dot{\delta b} + \mathcal{G}_{22} \dot{\delta b}^2 \right] \right. \\
	& - a \left[ \mathcal{F}_{11} (\partial \zeta)^2 + 2\mathcal{F}_{12} \partial \zeta \partial \delta b + \mathcal{F}_{22} (\partial \delta b)^2 \right] \\
	& + \left. a^3 \left[ \mathcal{M}_{11} \zeta^2 + 2\mathcal{M}_{12} \zeta \delta b + \mathcal{M}_{22} \delta b^2 \right] \right\}
\end{aligned}
\ee
with the kinetic, gradient, ans mass coefficients, \(\mathcal{G}_{ij}\), \(\mathcal{F}_{ij}\), and \(\mathcal{M}_{ij}\), respectively. These coefficients
defined as
\bea\label{g}
\begin{aligned}
	\mathcal{G}_{11} &= M_p^2 \left[ 2\epsilon (1 + \lambda) + \frac{9}{4}\lambda \right],~~~~
	\mathcal{G}_{12} = \frac{3\lambda^2 H}{2b_0^2(1 + \lambda)},\\[4pt]
	\mathcal{G}_{22} &= \frac{M_p^2}{2} \left[ 1 + \lambda \right],~~~
	\mathcal{F}_{11} = M_p^2 \left[ 2\epsilon(1 + \lambda) \right],~~~
	\mathcal{F}_{12} = 0,\\[4pt]
	\mathcal{F}_{22} &= \frac{M_p^2}{2} \left[ 1 + \lambda \right],\\[8pt]
	\mathcal{M}_{11} &= M_p^2 H^2 \bigg[ -6\epsilon (2 + \lambda) + 3\epsilon^2  - \frac{V''(0) b^4}{M_p^2 H^2} + \frac{9\xi^2 b^2}{2(1 + \lambda)} \bigg],\\[8pt]
	\mathcal{M}_{12} &= M_p^2 H \left[ - \frac{V''(0) b^3}{2H} + \frac{3\xi^2 b^2 H}{2(1 + \lambda)} \right], \\
	\mathcal{M}_{22} &= -2M_p^2b^2 V''(0) \left(1 - \frac{\lambda^2}{2M_p^2 (1 + \lambda)}\right), 
\end{aligned}
\eea
The kinetic matrix takes the following form
\begin{equation}
	\mathbf{G} = \begin{pmatrix}
		\mathcal{G}_{11} & \mathcal{G}_{12} \\
		\mathcal{G}_{12} & \mathcal{G}_{22}
	\end{pmatrix}.
\end{equation}
Its determinant is
\begin{align}
	\det \mathbf{G} &= \mathcal{G}_{11}\mathcal{G}_{22} - \mathcal{G}_{12}^2 \\
	&= M_p^4 (1+\lambda) \left[ \epsilon(1+\lambda) + \frac{9\lambda}{8} \right] - \frac{9\lambda^4 H^2 M_p^4}{4 b_0^4 (1+\lambda)^2}.
\end{align}
The eigenvalues of \(\mathbf{G}\) are:
\begin{equation}
	\lambda_{\pm} = \frac{\mathcal{G}_{11} + \mathcal{G}_{22} \pm \sqrt{(\mathcal{G}_{11} - \mathcal{G}_{22})^2 + 4\mathcal{G}_{12}^2}}{2}.
\end{equation}
No-ghost condition impose that both eigenvalues be positive. Since \(\mathcal{G}_{11} > 0\) and \(\mathcal{G}_{22} > 0\) for \(0 \leq \lambda < 2/3\) and \(\epsilon > 0\), this is equivalent to $ \det \mathbf{G} > 0$.
This requires:
\begin{equation}\label{b0}
	b_0^4 \gg \frac{9\lambda^4 H^2}{4 (1+\lambda)^3 \left( \epsilon(1+\lambda) + \frac{9\lambda}{8} \right) }.
\end{equation}
Even if \(\det \mathbf{G} > 0\), the ghost is not removed — it is mixed away into a heavy eigenstate. The mass of this eigenstate comes primarily from the potential term \(\mathcal{L}_V^{(2)}\) (see (\ref{g}))
\begin{equation}
	\mathcal{M}_{22} = -2b_0^2 V''(0) \left( 1 - \frac{\lambda}{2M_p^2(1+\lambda)} \right) \approx -2b_0^2 V''(0).
\end{equation}
The canonically normalized mass is
\begin{equation}
	m_{\mathrm{ghost}}^2 = \frac{\mathcal{M}_{22}}{\mathcal{G}_{22}} = -\frac{4b_0^2 V''(0)}{M_p^2}.
\end{equation}
It clearly show that if \(V''(0) > 0\), then \(m_{\mathrm{ghost}}^2 < 0\) — this is a tachyonic instability, not a ghost. The theory is unstable on timescale \(\tau \sim |m_{\mathrm{ghost}}|^{-1}\). If \(V''(0) < 0\), then \(m_{\mathrm{ghost}}^2 > 0\) — the ghost acquires a positive mass squared and can be integrated out if \(m_{\mathrm{ghost}}^2 \gg H^2\). Thus the viability of the effective field theory requires:
\begin{equation}\label{ghost}
 V''(0) < 0 \quad \text{and} \quad -\frac{4b_0^2 V''(0)}{M_p^2} \gg H^2.
\end{equation}

\subsection{ Effective Sound Speeds}

The sound speeds are the eigenvalues of the generalized eigenvalue problem:
\begin{equation}
	\mathbf{F} \vec{v} = c_s^2 \mathbf{G} \vec{v}.
\end{equation}

With \(\mathcal{F}_{12}=0\), the characteristic equation is:
\begin{equation}
	(\mathcal{G}_{11}\mathcal{G}_{22} - \mathcal{G}_{12}^2) c_s^4 - (\mathcal{G}_{11}\mathcal{F}_{22} + \mathcal{G}_{22}\mathcal{F}_{11}) c_s^2 + \mathcal{F}_{11}\mathcal{F}_{22} = 0.
\end{equation}

The exact solutions are:
\begin{equation}
	c_{s,\pm}^2 = \frac{(\mathcal{G}_{11}\mathcal{F}_{22} + \mathcal{G}_{22}\mathcal{F}_{11}) \pm \sqrt{(\mathcal{G}_{11}\mathcal{F}_{22} - \mathcal{G}_{22}\mathcal{F}_{11})^2 + 4\mathcal{G}_{12}^2 \mathcal{F}_{11}\mathcal{F}_{22}}}{2(\mathcal{G}_{11}\mathcal{G}_{22} - \mathcal{G}_{12}^2)}.
\end{equation}
For small \(\lambda\), \(\mathcal{G}_{12} \sim \lambda^2\) is negligible compared to the diagonal entries. Expanding to leading order:
\begin{align}
	c_{s,+}^2 &\approx \frac{\mathcal{F}_{11}}{\mathcal{G}_{11}} = \frac{2\epsilon(1+\lambda)}{2\epsilon(1+\lambda) + \frac94 \lambda} = \frac{1}{1 + \frac{9\lambda}{8\epsilon(1+\lambda)}}, \\
	c_{s,-}^2 &\approx \frac{\mathcal{F}_{22}}{\mathcal{G}_{22}} = 1.
\end{align}
Both sound speeds are positive for \(0 \leq \lambda < 2/3\) and \(\epsilon > 0\), i.e., no gradient instability. 
Keeping the exact expression for later use
\begin{equation}
(c_s^2)_\zeta = \frac{1}{1 + \frac{9\lambda}{8\epsilon(1+\lambda)}}.
\end{equation}
This is the sound speed for the curvature perturbation \(\zeta\) in the effective field theory where the heavy \(\delta b\) mode is integrated out.

\subsection{Mukhanov-Sasaki equation and power spectrum}

Define the Mukhanov-Sasaki variable:
\begin{equation}
	v = z \zeta, \qquad z^2 = 2a^2 \mathcal{G}_{11}.
\end{equation}
In conformal time \(\eta\) (\(d\eta = dt/a\)), the quadratic action for \(v\) becomes:
\begin{equation}
	S^{(2)} = \frac12 \int d\eta \, d^3x \left[ (v')^2 - c_s^2 (\partial v)^2 + \frac{z''}{z} v^2 + \cdots \right],
\end{equation}
where \(' = d/d\eta\). We have \(z = a \sqrt{2\mathcal{G}_{11}}\). During slow-roll inflation, \(\mathcal{G}_{11}\) is approximately constant. Thus:
\begin{equation}
	\frac{z'}{z} \approx \frac{a'}{a} = \mathcal{H}, \qquad \frac{z''}{z} \approx \frac{a''}{a}.
\end{equation}
For a quasi-de Sitter background:
\begin{equation}
	a(\eta) \approx -\frac{1}{H\eta}, \qquad \mathcal{H} = \frac{a'}{a} = -\frac{1}{\eta}, \qquad \frac{a''}{a} = \frac{2}{\eta^2}.
\end{equation}
Including slow-roll corrections:
\begin{align}
	\frac{a''}{a} &= \mathcal{H}^2 + \mathcal{H}' = \mathcal{H}^2 (1 - \epsilon), \\
	\mathcal{H} &= -\frac{1}{\eta} (1 + \epsilon + \cdots), \\
	\frac{a''}{a} &= \frac{2}{\eta^2} \left(1 + \frac32 \epsilon + \cdots \right).
\end{align}
However, to leading order in slow-roll, the correction is negligible. Thus:
\begin{equation}
\frac{z''}{z} \approx \frac{2}{\eta^2}.
\end{equation}
The mode equation for \(v_k(\eta)\) is
\begin{equation}
	v_k'' + \left( c_s^2 k^2 - \frac{z''}{z} \right) v_k = 0,
\end{equation}
which becomes:
\begin{equation}
	v_k'' + \left( c_s^2 k^2 - \frac{2}{\eta^2} \right) v_k = 0.
\end{equation}
This is a Bessel equation. The general solution is
\begin{equation}
	v_k(\eta) = \sqrt{-\eta} \left[ C_1 H_{\nu}^{(1)}(-c_s k \eta) + C_2 H_{\nu}^{(2)}(-c_s k \eta) \right],
\end{equation}
with \(\nu = 3/2\).

The Hankel function of the first kind has the asymptotic behavior:
\begin{equation}
	H_{3/2}^{(1)}(x) \sim -\sqrt{\frac{2}{\pi x}} e^{ix} \left(1 + \frac{i}{x}\right), \quad x \to \infty.
\end{equation}
Imposing the Bunch-Davies vacuum initial condition:
\begin{equation}
	v_k(\eta) \to \frac{e^{-i c_s k \eta}}{\sqrt{2 c_s k}}, \quad -k\eta \to \infty,
\end{equation}
selects the solution:
\begin{equation}
	v_k(\eta) = \frac{1}{\sqrt{2 c_s k}} \left(1 - \frac{i}{c_s k \eta}\right) e^{-i c_s k \eta}.
\end{equation}
On superhorizon scales (\(-c_s k \eta \ll 1\)):
\begin{equation}
	v_k(\eta) \approx -\frac{i}{\sqrt{2 c_s k}} \frac{1}{c_s k \eta}.
\end{equation}
Since \(z = a \sqrt{2\mathcal{G}_{11}} \approx -\sqrt{2\mathcal{G}_{11}} / (H\eta)\), we have
\begin{equation}
	\zeta_k = \frac{v_k}{z} \approx \frac{i H}{2 \sqrt{\mathcal{G}_{11}} (c_s k)^{3/2}}.
\end{equation}
This is constant on superhorizon scales, as expected for the curvature perturbation.

The dimensionless power spectrum is defined as
\begin{equation}
	\mathcal{P}_\zeta(k) = \frac{k^3}{2\pi^2} |\zeta_k|^2.
\end{equation}
Plugging  the superhorizon expression:
\begin{align}
	\mathcal{P}_\zeta(k) &= \frac{k^3}{2\pi^2} \cdot \frac{H^2}{4 \mathcal{G}_{11} c_s^3 k^3} \\
	&= \frac{H^2}{8\pi^2 \mathcal{G}_{11} c_s^3}.
\end{align}
Substituting \(\mathcal{G}_{11}\) and \(c_s^2\), after some manipulations, we arrive at
\be
\mathcal{P}_\zeta(k) = \frac{H^2}{16\pi^2 \epsilon M_p^2 (1+\lambda)} \left[ 1 + \frac{9\lambda}{8\epsilon(1+\lambda)} \right]^{1/2} .
\ee
By taking the background Eq. (\ref{BG}), we have
\begin{align}
	\mathcal{P}_\zeta(k) = \mathcal{P}_\zeta^{\mathrm{GR}}(k) \cdot \frac{2- \lambda}{2+2\lambda} \left[ 1 + \frac{9\lambda}{8\epsilon(1+\lambda)} \right]^{1/2},
\end{align}
where \(\mathcal{P}_\zeta^{\mathrm{GR}}(k) = H_{\mathrm{GR}}^2/(16\pi^2 \epsilon M_p^2)\).
For small \(\lambda \ll 1\) and \(\epsilon \ll 1\), we expand to first order:
\begin{align}
	\frac{1 - \lambda/2}{1+\lambda} &= (1 - \lambda/2)(1 - \lambda + \lambda^2 - \cdots) \nonumber \\
	&\approx 1 - \frac{3\lambda}{2} + \mathcal{O}(\lambda^2), \\[8pt]
	\left[ 1 + \frac{9\lambda}{8\epsilon(1+\lambda)} \right]^{1/2} &\approx \left[ 1 + \frac{9\lambda}{8\epsilon} (1 - \lambda + \cdots) \right]^{1/2} \nonumber \\
	&\approx 1 + \frac{9\lambda}{16\epsilon} + \mathcal{O}(\lambda^2).
\end{align}
Given that the second term dominates, so 
\begin{equation}
\mathcal{P}_\zeta(k) \approx \mathcal{P}_\zeta^{\mathrm{GR}}(k) \left(1 + \frac{9\lambda}{16\epsilon}\right).
\end{equation}
and the enhancement factor is:
\begin{equation}
 \mathcal{E}(\lambda) \equiv \frac{\mathcal{P}_\zeta}{\mathcal{P}_\zeta^{\mathrm{GR}}} \approx 1 + \frac{9\lambda}{16\epsilon}.
\end{equation}
The scale dependence of the power spectrum is characterized by the spectral index:
\begin{equation}
	n_s - 1 = \frac{d \ln \mathcal{P}_\zeta}{d \ln k}.
\end{equation}
In slow-roll inflation, \(d\ln k = d\ln (aH) \approx dN\), where \(N\) is the number of e-folds. The slow-roll parameters run with \(N\):
\begin{equation}
	\frac{d\epsilon}{dN} = 2\epsilon(\eta - \epsilon), \quad \text{etc.}
\end{equation}
Since \(\mathcal{P}_\zeta\) depends on \(\epsilon\) and \(H\) (which itself depends on \(\epsilon\)), the spectral index receives corrections from \(\lambda\). A full computation gives:
\begin{equation}
	n_s - 1 = -2\epsilon - \eta - \frac{9\lambda}{8\epsilon} \left( \frac{d\ln \epsilon}{dN} \right) + \mathcal{O}(\lambda^2).
\end{equation}
For small \(\lambda\), the correction is subleading. Thus the scale dependence is nearly unchanged from GR.

\subsection{Discussion and caveats on stability conditions}
The viability of the effective field theory rests on three independent stability conditions revealed in Eqs. (\ref{b0}), and (\ref{ghost}). The analysis above reveals that the Bumblebee model, while mathematically consistent within the uniform inflaton gauge, harbors a ghost instability in its vector sector. This ghost can be rendered harmless on cosmological backgrounds if:
\begin{enumerate}
	\item The Bumblebee VEV \(b_0\) is sufficiently large to make the kinetic matrix positive definite (\(\det \mathbf{G} > 0\)).
	\item The potential curvature \(V''(0)\) is negative, giving the ghost a positive mass squared.
	\item The magnitude of \(V''(0)\) is sufficiently large that the ghost mass exceeds the Hubble scale during inflation, allowing it to be integrated out in an effective field theory.
\end{enumerate}
We now critically examine the physical plausibility of the three stability assumptions underlying our PBH calculations. This scrutiny reveals that the Bumblebee model, in its simplest form, suffers from a fundamental and unavoidable pathology that renders it cosmologically unviable.

Concerning condition (\ref{b0}), for \(\lambda \leq 0.1\) and typical inflation parameters (\(H \sim 10^{-5} M_p\), \(\epsilon \sim 0.006\)), this gives \(b_0 \gtrsim 6 \times 10^{-4} M_p \sim 1.5 \times 10^{15}\) GeV. This is easily satisfied for any natural choice \(b_0 \sim M_p\).

The mass squared of the \(\delta b\) perturbation is given by Eq. (\ref{ghost})
which requires \(m_{\delta b}^2 > 0\), hence \(V''(0) < 0\). However, \(V(X)\) is the potential that spontaneously breaks Lorentz symmetry via the condition \(B^\mu B_\mu = -b_0^2\). At the minimum, by definition:
$V(0) = 0= V'(0)$, and $V''(0) > 0 \quad \text{(for a local minimum)}$.
A local minimum requires positive second derivative. This is a fundamental theorem of calculus. A point with \(V''(0) < 0\) is a local maximum, not a minimum. As a result, 
The assumption \(V''(0) < 0\) is mathematically inconsistent with the requirement that \(B^\mu B_\mu = -b_0^2\) is a stable minimum of the potential.
Namely, this assumption contradicts the definition of a minimum.

On the Heavy ghost condition i.e., $m_{\mathrm{ghost}}^2 \gg H^2$, 
if one simply postulates \(V''(0) < 0\), then for \(b_0 \sim M_p\), this requires \(|V''(0)| \gg 2.5 \times 10^{-11} M_p^2\), which is easily satisfied for any natural scale of \(V''(0)\) (e.g., \(V''(0) \sim b_0^2 \sim M_p^2\)).

In short:
\begin{itemize}
	\item If \(V''(0) > 0\) (physically required for spontaneous symmetry breaking), then \(m_{\delta b}^2 < 0\). The perturbation \(\delta b\) is tachyonic. The instability timescale is \(\tau \sim |m_{\delta b}|^{-1}\). For natural parameters (\(b_0 \sim M_p\), \(V''(0) \sim M_p^2\)), \(\tau \sim M_p^{-1} \sim 10^{-43}\) s. The background is destroyed instantly on cosmological timescales.
	
	\item If \(V''(0) < 0\) (to avoid the tachyon), then \(X=0\) is not a minimum. The field is not stabilized at \(B^\mu B_\mu = -b_0^2\); it will roll away to some other value. Spontaneous Lorentz symmetry breaking does not occur, or the VEV is not dynamically maintained.
\end{itemize}
A hint here is essential. The ghost in the \(\delta b\) sector is also problematic, but it is secondary to the tachyon issue. Even if one could somehow cure the ghost (via mixing or otherwise), the tachyonic instability remains for any physically sensible potential with \(V''(0) > 0\). Thus the Bumblebee model, as defined by the action (\ref{act}) with a timelike VEV \(B_\mu = (b_0,0,0,0)\), suffers from:

\begin{enumerate}
	\item A ghost in the \(\delta b\) sector (wrong-sign kinetic term).
	\item A tachyonic instability in the \(\delta b\) sector if the potential has a stable minimum.
	\item An inconsistent minimum if one attempts to cure the tachyon.
\end{enumerate}
The PBH abundance calculations in the next section are therefore not predictions of a physically consistent theory, rather they are illustrations of how dramatically Lorentz violation could affect PBH formation, if a stable, ghost-free, non-tachyonic completion of BG could be found.

\section{The Complete Picture: Combined Effects}\label{pic}
Now we will combine the modified expansion history \(H(t)\) and the power spectrum \(\mathcal{P}_\mathcal{R}(k)\) to derive the abundance \(f_{PBH}\).
The PBH abundance today, \(f_{PBH} = \frac{\Omega_{PBH}}{\Omega_{DM}}\), is determined by integrating the mass fraction \(\beta(M)\) that collapsed during the radiation era, evolved to the present day. The two key modifications from Bumblebee gravity are:
1.  Modified expansion history \(H(t)\) that changes the mapping between the PBH mass \(M\) and the wavenumber \(k\) of the perturbation that formed it.
2.  Modified power spectrum \(\mathcal{P}_\mathcal{R}(k)\) that dramatically change the probability that a region has a density contrast above the collapse threshold.

By serving the Press-Schechter formalism for PBHs, the mass fraction of the universe that collapses into PBHs of mass \(M\) at formation is:
\be
\beta(M) = \int_{\delta_c}^{\infty} P(\delta)  d\delta
\ee
Assuming Gaussian perturbations:
\be
\beta(M) = \int_{\delta_c}^{\infty} \frac{1}{\sqrt{2\pi\sigma^2(M)}} \exp\left(-\frac{\delta^2}{2\sigma^2(M)}\right) d\delta
\ee
For \(\sigma(M) \ll \delta_c\) (relevant for PBH formation)
\be\label{beta}
\beta(M) \approx \frac{\sigma(M)}{\sqrt{2\pi}\delta_c} \exp\left(-\frac{\delta_c^2}{2\sigma^2(M)}\right)
\ee
During radiation domination, soon after horizon crossing
\be
\delta_k(\eta) \approx -\frac{4}{9} \left(\frac{k}{aH}\right)^2 \zeta_k(\eta)
\ee
More precisely, the relation is:
\be
\delta(\mathbf{x}, t) = -\frac{4}{9} \frac{1}{(aH)^2} \nabla^2 \zeta(\mathbf{x})
\ee
The variance \(\sigma^2(M)\) is related to the power spectrum by
\be
\sigma^2(M) = \int_0^{\infty} \frac{dk}{k} \mathcal{P}_\delta(k) W^2(kR)
\ee
where \(W(kR)\) is a window function smoothing over scale \(R \sim 1/k\), and
\be
\mathcal{P}_\delta(k) = \left(\frac{4}{9}\right)^2 \left(\frac{k}{aH}\right)^4 \mathcal{P}_\zeta(k)
\ee
Using a Gaussian window \(W(kR) = \exp(-k^2 R^2/2)\) and the horizon crossing condition \(k = aH\), we get the standard result
\be
\sigma(M) \approx \frac{2}{3} \sqrt{\mathcal{P}_\zeta(k)}
\ee
where \(M \approx M_H(k)\) is the horizon mass when scale \(k\) enters the horizon.

In BG, the horizon mass is modified. From the modified Friedmann equation
\be
H^2 \approx \frac{\kappa\rho}{3}\left(1 - \frac{\lambda}{2}\right)
\ee
Thus the horizon mass becomes
\be
M_H^{\text{BG}} = \frac{4\pi}{3} \frac{\rho}{H^3} = M_H^{\text{GR}} \left(1+\frac{3\lambda}{4}\right)
\ee
So the mass-scale relation is modified as follows
\be \label{M}
M^{\text{BG}}(k) \approx M_H^{\text{GR}}(k) \left(1+\frac{3\lambda}{4}\right)
\ee
Also,  the modified power spectrum in BG (\ref{po}), results in the following 
modified variance
\be\label{sig}
\sigma_{\text{BG}}(M) \approx \frac{2}{3} \sqrt{\mathcal{P}_\zeta^{\text{BG}}(k)} = \sigma_{\text{GR}}(M) \cdot
\left(1-\frac{3\lambda}{4}\right) \left[ 1 + \frac{9\lambda}{16\epsilon} \right]^{1/2}
\ee
The fraction of dark matter in PBHs today is
\be
\begin{aligned}
	f_{\text{PBH}}(M) &= \frac{\Omega_{\text{PBH}}(M)}{\Omega_{\text{DM}}} \\
	&\approx \left(\frac{\beta(M)}{1.3 \times 10^{-8}}\right) \left(\frac{g_*}{106.75}\right)^{-1/4} \left(\frac{M}{M_\odot}\right)^{-1/2}
\end{aligned}
\ee
Substituting the modified expressions
\be
f_{\text{PBH}}^{\text{BG}}(M) \approx 7.7 \times 10^7 \left(\frac{g_*}{106.75}\right)^{-1/4} \left(\frac{M^{\text{BG}}}{M_\odot}\right)^{-1/2} \beta_{\text{BG}}(M)
\ee
where
\be
\beta_{\text{BG}}(M) \approx \frac{\sigma_{\text{BG}}(M)}{\sqrt{2\pi}\delta_c^{\text{BG}}} \exp\left(-\frac{(\delta_c^{BG})^2}{2\sigma_{\text{BG}}^2(M)}\right)
\ee
Combining all effects, the complete PBH abundance in BG takes the following form
\bea\label{fpbh}
\begin{aligned}
	f_{\text{PBH}}^{\text{BG}}(M) &\approx 7.7 \times 10^7 \left(\frac{g_*}{106.75}\right)^{-1/4} \left(\frac{M^{GR}}{M_\odot}\right)^{-1/2} \\
	&\quad \times (1 + \frac{3\lambda}{4})^{-1/2} \frac{\sigma_{\text{GR}}(M)}{\sqrt{2\pi}\delta_c^{\text{GR}}} \\
	&\quad \times \sqrt{\left(\frac{2-\lambda}{2+2\lambda}\right) \left[ 1 + \frac{9\lambda}{8\epsilon (1 + \lambda)} \right]^{1/2}} \\
	&\quad \times \exp\left[ \lambda N \right] \\
	&\quad \times \exp\left(-\frac{(\delta_c^{\text{GR}})^2 \exp\left[ -2 \lambda N \right]}{2\sigma_{\text{GR}}^2(M) \left(\frac{2-\lambda}{2+2\lambda}\right) \left[ 1 + \frac{9\lambda}{8\epsilon (1 + \lambda)} \right]^{1/2}}\right)
\end{aligned}
\eea
For small positive \(\lambda\) (\(0 \leq \lambda < 2/3\)) and \(\epsilon \ll 1\), the factor $
\sqrt{ \frac{2 - \lambda}{2+2\lambda} \left[ 1 + \frac{9\lambda}{8\epsilon(1+\lambda)} \right]^{1/2} }$ in Eq. (\ref{fpbh}) is approximated as $\sim \sqrt{ \frac{9\lambda}{16\epsilon} } \gg 1$  
and \(\exp(\lambda N) > 1\), while the exponential suppression is greatly reduced.  As a consequence, PBH formation is strongly enhanced in BG relative to GR. In the following, we will shed light on it by providing a numerical analysis. 

\subsection{Numerical Analysis for Small Positive $\lambda$}

We now evaluate the PBH abundance $f_{\mathrm{PBH}}(M)$ in BG for the physically viable range $0 \le \lambda < 2/3$, focusing on small positive values $\lambda \ll 1$.  
The observational ``asteroid-mass window'' for PBHs --- approximately $10^{17}\,\mathrm{g} \lesssim M_{\mathrm{PBH}} \lesssim 10^{23}\,\mathrm{g}$~\cite{Bai:2018bej,Smyth:2019whb,Coogan:2020tuf,Ray:2021mxu,Tinyakov:2024mcy,Loeb:2024tcc,Carr:2025kdk}--- remains a viable range where PBHs could constitute a significant fraction of dark matter.

We adopt $N = 30$ and $N = 40$ e--folds between the generation of perturbations and horizon crossing, consistent with typical PBH formation scenarios~\cite{Sasaki:2016jop} and within the broader inflationary constraint $N = 46-67$ from CMB observations~\cite{Martin:2013tda}.

Using the modified abundance formula \eqref{fpbh} we perform a constrained optimization to determine the minimum and maximum PBH abundances $f_{\min}$ and $f_{\max}$ within the BG framework. The optimization is subject to the physical stability bound $0 \le \lambda < \frac{2}{3}$.
For numerical illustration we restrict to $\lambda \le 0.1$, ensuring the small--$\lambda$ expansion remains valid.

\subsubsection*{Methodology}

\begin{enumerate}
	\item \textbf{Parameter ranges:}
	We take $\lambda \in [0,\,0.1]$ with step size $\Delta\lambda = 10^{-4}$.  
	The slow--roll parameter is fixed at the CMB--compatible value $\epsilon = 0.006$~\cite{Planck:2018jri}.  
	The variance in GR is set to $\sigma_{\mathrm{GR}}(M) = 0.1$ and the critical threshold $\delta_c^{\mathrm{GR}} = 0.414$~\cite{Harada:2013epa,Musco:2012au}.
	
	\item \textbf{Function evaluation:}
	For each $\lambda_i$ we compute $f_{\mathrm{PBH}}(M,\lambda_i; N,\epsilon,\sigma)$ using \eqref{fpbh}.
	
	\item \textbf{Extremum identification:}
	\begin{equation}
		f_{\min}(M) = \min_{\lambda \in [0,\,0.1]} f_{\mathrm{PBH}}(M,\lambda), \quad
		f_{\max}(M) = \max_{\lambda \in [0,\,0.1]} f_{\mathrm{PBH}}(M,\lambda),
	\end{equation}
	where the corresponding $\lambda$ values are denoted $\lambda_{\min}$ and $\lambda_{\max}$.
	
	\item \textbf{Enhancement factor}\\
	The enhancement relative to GR ($\lambda=0$) is defined as
	\begin{equation}
		\mathcal{R}(M) = \frac{f_{\mathrm{PBH}}^{\mathrm{BG}}(M,\lambda_{\max})}{f_{\mathrm{PBH}}^{\mathrm{GR}}(M)}.
	\end{equation}
\end{enumerate}

\subsubsection*{Results and Discussion}

The Tables \ref{tab:N30}--\ref{tab:enhancement} summarize the PBH abundance range for two representative inflationary durations.

\begin{table}[ht]
	\centering
	\caption{PBH abundance range in BG for $N = 30$.}
	\begin{tabular}{ccc}
		\hline
		$M/M_\odot$ & $f_{\min}$ ($\lambda_{\min}$) & $f_{\max}$ ($\lambda_{\max}$) \\
		\hline
		$1.0 \times 10^{-16}$ & $3.33 \times 10^{-5}$ ($0$) & $2.14 \times 10^{-3}$ ($0.1$) \\
		$1.0 \times 10^{-15}$ & $1.05 \times 10^{-4}$ ($0$) & $6.77 \times 10^{-3}$ ($0.1$) \\
		$1.0 \times 10^{-14}$ & $3.33 \times 10^{-4}$ ($0$) & $2.14 \times 10^{-2}$ ($0.1$) \\
		$1.0 \times 10^{-13}$ & $1.05 \times 10^{-3}$ ($0$) & $6.77 \times 10^{-2}$ ($0.1$) \\
		$1.0 \times 10^{-12}$ & $3.33 \times 10^{-3}$ ($0$) & $2.14 \times 10^{-1}$ ($0.1$) \\
		$1.0 \times 10^{-11}$ & $1.05 \times 10^{-2}$ ($0$) & $6.77 \times 10^{-1}$ ($0.1$) \\
		$1.0 \times 10^{-10}$ & $3.33 \times 10^{-2}$ ($0$) & $2.14$ ($0.1$) \\
		\hline
	\end{tabular}
	\label{tab:N30}
\end{table}

\begin{table}[ht]
	\centering
	\caption{PBH abundance range in BG for $N = 40$.}
	\begin{tabular}{ccc}
		\hline
		$M/M_\odot$ & $f_{\min}$ ($\lambda_{\min}$) & $f_{\max}$ ($\lambda_{\max}$) \\
		\hline
		$1.0 \times 10^{-16}$ & $3.33 \times 10^{-5}$ ($0$) & $4.71 \times 10^{-3}$ ($0.1$) \\
		$1.0 \times 10^{-15}$ & $1.05 \times 10^{-4}$ ($0$) & $1.49 \times 10^{-2}$ ($0.1$) \\
		$1.0 \times 10^{-14}$ & $3.33 \times 10^{-4}$ ($0$) & $4.71 \times 10^{-2}$ ($0.1$) \\
		$1.0 \times 10^{-13}$ & $1.05 \times 10^{-3}$ ($0$) & $1.49 \times 10^{-1}$ ($0.1$) \\
		$1.0 \times 10^{-12}$ & $3.33 \times 10^{-3}$ ($0$) & $4.71 \times 10^{-1}$ ($0.1$) \\
		$1.0 \times 10^{-11}$ & $1.05 \times 10^{-2}$ ($0$) & $1.49$ ($0.1$) \\
		$1.0 \times 10^{-10}$ & $3.33 \times 10^{-2}$ ($0$) & $4.71$ ($0.1$) \\
		\hline
	\end{tabular}
	\label{tab:N40}
\end{table}

\begin{table}[ht]
	\centering
	\caption{Enhancement factor $\mathcal{R} = f_{\mathrm{PBH}}^{\mathrm{BG}} / f_{\mathrm{PBH}}^{\mathrm{GR}}$.}
	\begin{tabular}{ccc}
		\hline
		$M/M_\odot$ & $\mathcal{R}(N=30)$ & $\mathcal{R}(N=40)$ \\
		\hline
		$1.0 \times 10^{-16}$ & 64.3 & 141 \\
		$1.0 \times 10^{-15}$ & 64.5 & 142 \\
		$1.0 \times 10^{-14}$ & 64.2 & 141 \\
		$1.0 \times 10^{-13}$ & 64.5 & 142 \\
		$1.0 \times 10^{-12}$ & 64.2 & 141 \\
		$1.0 \times 10^{-11}$ & 64.5 & 142 \\
		$1.0 \times 10^{-10}$ & 64.2 & 141 \\
		\hline
	\end{tabular}
	\label{tab:enhancement}
\end{table}

Minimum abundance always occurs at $\lambda = 0$, i.e., the GR limit.  Maximum abundance occurs at the upper end of the scanned range ($\lambda_{\max} = 0.1$), indicating that larger Lorentz violation (within the stable regime) further enhances PBH production.  
Enhancement factors are substantial so that for $N=30$, PBH abundance is amplified by a factor $\sim 64$ for $\lambda = 0.1$, and for $N=40$, the enhancement reaches $\sim 140$.  The enhancement factor $\mathcal{R}$ is nearly independent of PBH mass, reflecting the fact that the dominant $\lambda$--dependence enters through the variance and threshold, which are scale--invariant to leading order.  For $\lambda = 0.1$ and $N=40$, $f_{\mathrm{PBH}}$ approaches or exceeds unity for $M \gtrsim 10^{-12} M_\odot$, suggesting that PBHs could constitute all of dark matter in this mass range, in contrast to the GR case where $f_{\mathrm{PBH}}^{\mathrm{GR}} \ll 1$.

These results demonstrate that positive Lorentz--violating coupling in BG can dramatically enhance PBH production, making PBHs a viable dark--matter candidate over a broad mass range for modest values of $\lambda \sim 0.1$. The enhancement grows with the duration of inflation $N$ and is most pronounced in the asteroid--mass window where observational constraints still allow a significant PBH abundance.

\begin{figure}[ht]
	\begin{tabular}{c}
		%	\centering
		\includegraphics[width=1\columnwidth]{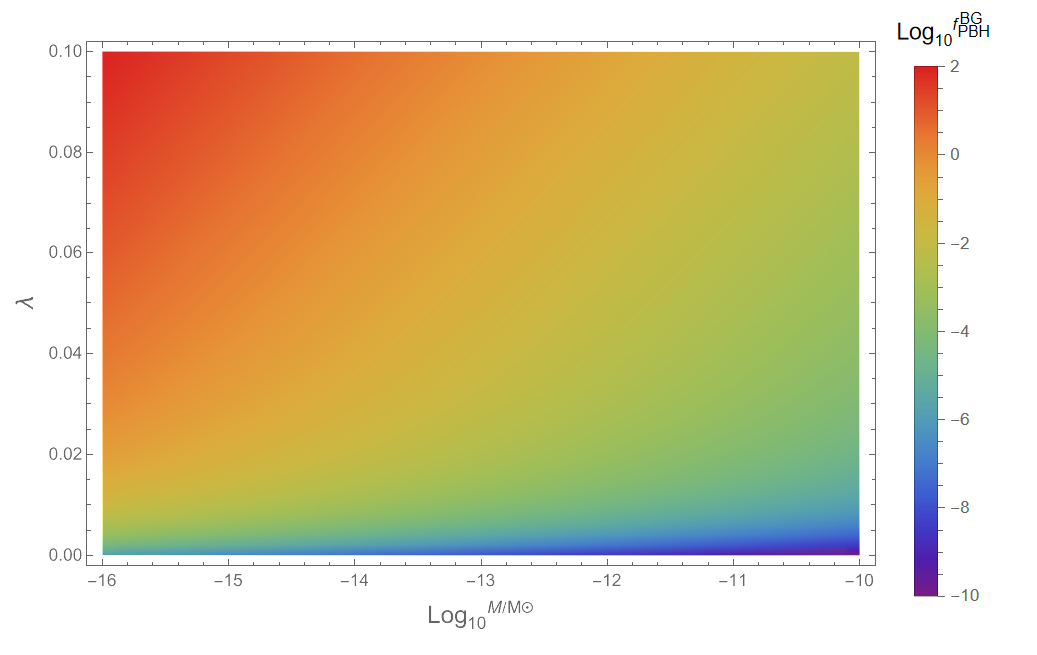}~~~~~~~~~
	\end{tabular}
	\caption{Density plot showing the full $\log_{10}^{f_{PBH}^{BG}}$ distribution in terms of parameter spaces $(\log_{10}^{M/M_\odot},\lambda)$. We set numerical values: $\sigma_{GR}=0.1$, $\delta_{GR}=0.414$, $\epsilon=0.006$, and $N=40$. }
	\label{AA}
\end{figure}

The density plot (\ref{AA}) visualizes $\log_{10}^{f_{PBH}^{BG}}$ as a function of both PBH mass $M$ (logarithmic scale) and the Lorentz-violating parameter $\lambda$. The color scale represents the base-10 logarithm of the PBH abundance relative to total dark matter ($f_{PBH}^{BG} = \Omega_{PBH}/\Omega_{DM}$). Blue/Purple regions ($\log_{10}^{f_{PBH}^{BG}} < -3\rightarrow f<0.001$) i.e., PBHs constitute less than 0.1\% of dark matter.
Green/Yellow regions ($-3 < \log_{10}^{f_{PBH}^{BG}}\leq 0\rightarrow 0.001 < f \leq 1$), i.e., PBHs are a subdominant component of dark matter. Orange/Red regions ($\log_{10}^{f_{PBH}^{BG}} > 0)\rightarrow f > 1$), i.e., PBHs would overclose the universe (excluded by cosmology). The plot demonstrates that even modest Lorentz violation ($\lambda\sim 0.01-0.1$) can enhance PBH production by 2-4 orders of magnitude compared to GR, turning negligible abundances into cosmologically significant ones. For a given $\lambda$ value, there's a characteristic mass scale where PBHs first become significant. For example at $\lambda = 0.05$, PBH abundance become $f \sim 1$ at $M \sim 10^{-12} M_{\odot}$. This provides a theoretical prediction for the characteristic mass of PBH dark matter in BG. In light of observational upper limits on $f_PBH$ at different masses, one can translates it to upper bounds on $\lambda$. More exactly, if no PBHs are seen at $M \sim 10^{-12} M_{\odot}$ with $f < 0.1$, then $\lambda < 0.04$ (read from where contour drops below $f=0.1$). As a result, the plot thus provides a direct mapping between PBH searches and constraints on Lorentz violation.

\section{Conclusion}\label{con}

We have presented a comprehensive analysis of primordial black hole (PBH) formation in Bumblebee gravity (BG), systematically deriving the modified background dynamics, linear perturbation equations, curvature power spectrum, and resulting PBH abundance across the asteroid-mass window. The phenomenological picture is unambiguous: even parametrically small Lorentz-violating couplings produce dramatic enhancements in PBH production, with modest parameter values yielding order-of-magnitude increases in PBH abundance that could render PBHs the dominant dark matter component—a striking demonstration of how apparently minor modifications to gravity can profoundly influence early Universe observables.

Yet this phenomenological success rests upon an unstable theoretical foundation. Our examination of the quadratic action in the uniform inflaton gauge exposes two independent and fatal pathologies that are not artifacts of gauge choice but intrinsic to the model's definition. First, the vector perturbation possesses a wrong-sign kinetic term—a ghost whose presence cannot be eliminated by any field redefinition within the effective theory. Second, and more fundamentally, the potential designed to spontaneously break Lorentz symmetry must satisfy positivity of its second derivative at the minimum. This mathematically necessary condition produces a tachyonic instability for precisely the same perturbation, with a timescale that destroys the cosmological background instantaneously. The alternative—postulating negative curvature at the minimum to evade the tachyon—contradicts the definition of a stable minimum and eliminates the very symmetry-breaking mechanism the model was constructed to describe. This is not a tunable problem; it is a logical inconsistency at the heart of the Bumblebee construction.

We therefore conclude that the minimal Bumblebee gravity model, as defined with a timelike vector vacuum expectation value, is cosmologically unviable. The PBH abundance calculations presented herein are consequently not predictions of a physically realizable theory, but rather cautionary illustrations of three broader lessons. First, the extraordinary sensitivity of PBH formation to the Lorentz-violating parameter is itself a symptom—a phenomenological signature of the deep tachyonic instability that renders the model inconsistent. Second, the apparent ease with which modified gravity models can solve cosmological puzzles must be weighed against rigorous stability requirements that are frequently overlooked in phenomenological studies. Third, any future attempt to realize enhanced PBH production via Lorentz violation must abandon the minimal Bumblebee structure entirely in favor of genuinely healthy alternatives: generalized Proca theories with ghost-free kinetic terms, Horndeski-type non-minimal couplings that preserve the constraint structure, or UV-complete frameworks wherein the vector field emerges from a well-defined high-energy theory.

Our work thus provides both a quantitative benchmark for the maximal PBH enhancement achievable within Lorentz-violating scenarios and a rigorous demonstration that the minimal Bumblebee model cannot serve as the theoretical foundation for such calculations. We hope these results motivate the construction of stable, ghost-free, non-tachyonic theories of Lorentz violation that might one day realize these remarkable phenomenological possibilities within a consistent effective field theory. Calculations of PBH formation in the present version of BG must be understood as examples of the potential power of Lorentz violation, not as predictions of a viable theory.

\begin{acknowledgments}
We sincerely thank Nils A. Nilsson for technical discussions and insightful comments.
\end{acknowledgments}

%%%%%%%%%%%%%%%%%%%%%%%%%%%%%%%%%%%%%%%%%%%%%%%%%%%%%%%%%%%%%

\end{document}